\documentclass[preprint,12pt]{elsarticle}

\usepackage{amssymb}
\newcommand{\ff}{\frac{1}{2}}
\newcommand{\nn}{\nonumber}

\journal{Synthetic Metals}

\begin{document}

\begin{frontmatter}



\title{Onsager phase factor of quantum oscillations \\ in the organic metal $\theta$-(BEDT-TTF)$_4$CoBr$_4$(C$_6$H$_4$Cl$_2$)}


\author[1]{Alain~Audouard\fnref{fn1}\corref{cor1}}
\cortext[cor1]{Corresponding author}
\fntext[fn1]{Tel.: +33 562172869}
\ead{alain.audouard@lncmi.cnrs.fr}

\address[1]{Laboratoire National des Champs Magn\'{e}tiques
Intenses (UPR 3228 CNRS, INSA, UJF, UPS) 143 avenue de Rangueil,
F-31400 Toulouse, France.}

\author[2]{Jean-Yves~Fortin}
\ead{fortin@ijl.nancy-universite.fr}

\author[1]{David~Vignolles}

\address[2]{Institut Jean Lamour, D\'epartement de Physique de la
Mati\`ere et des Mat\'eriaux,
CNRS-UMR 7198, Vandoeuvre-les-Nancy, F-54506, France.}

\author[3]{Rustem~B.~Lyubovskii}

\address[3]{Institute of Problems of
Chemical Physics, RAS, 142432 Chernogolovka, MD, Russia.}

\author[3]{Elena~I.~Zhilyaeva}

\author[3]{Rimma~N.~Lyubovskaya}

\author[4]{Enric Canadell}

\address[4]{Institut de Ci\`{e}ncia de Materials de Barcelona, CSIC, Campus de la UAB, 08193, Bellaterra, Spain.}

\begin{abstract}
De Haas-van Alphen oscillations are studied for Fermi surfaces illustrating the Pippard's model, commonly observed
in multiband organic metals. Field- and temperature-dependent amplitude of the various Fourier components, linked
to frequency combinations arising from magnetic breakdown between different bands, are considered. Emphasis is
put on the Onsager phase factor of these components. It is demonstrated that, in addition to the usual Maslov index,
field-dependent phase factors must be considered to precisely account for the
data at high magnetic field.
We present compelling evidence of the existence of such contributions for the organic metal
$\theta$-(BEDT-TTF)$_4$CoBr$_4$(C$_6$H$_4$Cl$_2$).
\end{abstract}


\end{frontmatter}

% \linenumbers

\section{Introduction}
\label{}
Fermi surface (FS) of numerous organic metals is an illustration of the textbook model proposed by Pippard more than fifty years ago to compute Landau band structure induced by magnetic breakdown (MB) in multiband metals \cite{Pi62}. This is the case of the FS of the strongly two-dimensional charge transfer salt $\theta$-(BEDT-TTF)$_4$CoBr$_4$(C$_6$H$_4$Cl$_2$) (where BEDT-TTF stands for the bis-ethylenedithio-tetrathiafulvalene molecule), which is reported in Fig.~\ref{fig:FS_orbits} \cite{Au12}. Organic metals with such a FS are known to give rise to magnetic oscillations spectra involving linear combinations of the frequencies linked to the basic orbit $\alpha$ and the MB orbit $\beta$. These frequencies correspond not only to MB orbits such as $\beta+\alpha$  or harmonics but also to 'forbidden frequencies' such as $\beta-\alpha$ that are not predicted by the semiclassical model of Falicov-Stachowiak \cite{Fa66,Sh84}.

Only recently, analytic tools have been provided to account for the field and
temperature dependence of the Fourier amplitude relevant to the various frequencies observed \cite{Au12}. Besides, to our knowledge, little attention has been paid to the Onsager phase factor, yet. Though, according to Slutskin and Kadigrobov
\cite{Sl67} and Kochkin \cite{Ko68}, a field-dependent Onsager phase should be
observed for the considered FS, provided the magnetic field is large enough
compared to the MB field. Almost ten years later, the same result was
independently derived \cite{Hu76} in order to account for the discrepancy
between calculations, which are valid for the low field range, and the
experimental data for the lens orbit of Cd \cite{Co71} which share similarities
with the $\alpha$ orbit of Fig.~\ref{fig:FS_orbits}. However, still to our
knowledge, no further study in this field has been reported up to now. In order
to address this question, this paper is focused on the Onsager phase factor of
the various Fourier components observed in the de Haas-van Alphen (dHvA) spectrum of the organic metal $\theta$-(BEDT-TTF)$_4$CoBr$_4$(C$_6$H$_4$Cl$_2$) in fields of up to 55 T.

\section{Model}
\label{sec:model}

In this section, we first present the model accounting for the field and temperature dependence of the amplitude of the various Fourier components entering the oscillation spectra \cite{Au12}. In the second step, the field-dependent Onsager phase is considered.

\subsection{Fourier amplitude}

\begin{figure}
\centering
%\resizebox{\columnwidth}{!}
\resizebox{12cm}{!}{\includegraphics*{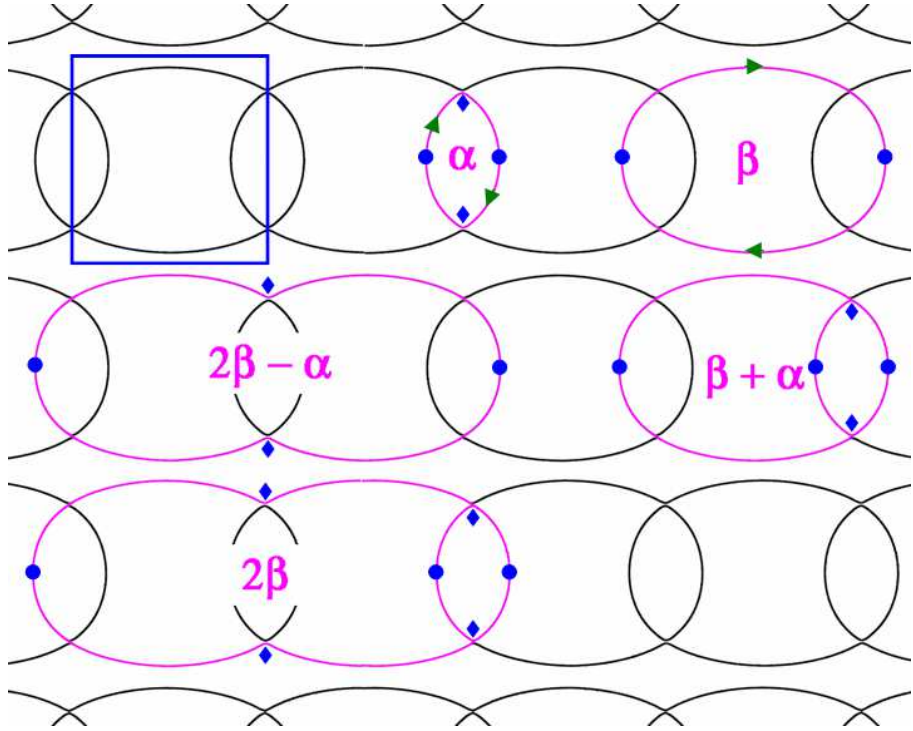}}
\caption{\label{fig:FS_orbits} (Color online) Fermi surface of $\theta$-(BEDT-TTF)$_4$CoBr$_4$(C$_6$H$_4$Cl$_2$) \cite{Au12} in the extended zone scheme.
Blue solid lines depict the first Brillouin zone. Pink lines display the
classical orbits considered for the data analysis, and arrows indicate the
quasi-particles path on the principal orbits $\alpha$ and $\beta$.
Blue circles and diamonds indicate the turning points in the direction parallel to the chains and the Bragg reflection points, respectively.}
\end{figure}

As displayed in Fig.~\ref{fig:FS_orbits}, the FS is composed of the $\alpha$ quasi-two-dimensional closed tube and a pair of quasi-one-dimensional sheets separated from the $\alpha$ orbit by a gap liable to be overcome by MB.  Numerous classical MB orbits can be defined ($\eta$ = $\alpha$, $\beta$, $\alpha+\beta$, 2$\beta-\alpha$, 2$\beta$, $etc.$), the area of which are linear combinations of those relevant to the  $\alpha$ and  $\beta$ orbits. The area of the latter is equal to that of the first Brillouin zone (FBZ). It can be remarked that 2$\beta$ corresponds to both the classical orbit displayed in Fig.~\ref{fig:FS_orbits} and to the 2$^{nd}$ harmonic of $\beta$.

To account for this FS, a two-band system with band extrema $\Delta_{0(1)}$ and effective masses $m_{0(1)}$ (in units of the electron mass $m_e$) is considered \cite{Au12}. The band 0 gives rise to the quasi-one-dimensional part of the FS of
Fig.~\ref{fig:FS_orbits} whereas the $\alpha$ orbit is built on the band 1. Assuming parabolic dispersion, the relevant frequency  is $F_{\alpha}=m_{1}(\mu-\Delta_{1})$. The $\beta$ orbit, generated by four tunnelings at the junction points, is built on both bands 0 and 1 and, still for a parabolic band, has a frequency corresponding to the first Brillouin zone area, $F_{\beta}=m_{\alpha}(\mu-\Delta_{\alpha})+m_{0}(\mu-\Delta_{0})=m_{\beta} (\mu-\Delta_{\beta}).$ In this case we identify the mass $m_0+m_{\alpha}$ with the mass $m_{\beta}$ of the orbit $\beta$.

To compute the oscillating part of the magnetization at fixed number $N$ of quasi-particles, we need to consider the oscillatory part of the free energy, defined by

\begin{eqnarray}
F(T,N,B)= \Omega(T,\mu,B)+N\mu
\end{eqnarray}

For a constant $N$, the oscillatory part of the grand potential $\Omega$ for a sample slab with area $\cal A$ can be written

\begin{eqnarray}
\nonumber
\phi_0\frac{u_0}{k_B}\frac{\Omega(T,\mu,B)}{\cal A}&=&
-\frac{m_0}{2}(\mu-\Delta_0)^2-\frac{m_{1}}{2}(\mu-\Delta_{1})^2
\\
\label{eq:omega}
&+&\frac{B^2}{2}\sum_{p\ge 1}\sum_{\eta}\frac{C_{\eta}}{\pi^2p^2m_{\eta}}
R_{\eta,p}(T)\cos(2\pi pF_{\eta}/B+p\varphi_{\eta}).
\end{eqnarray}

Damping factors can be expressed as $R_{\eta,p}(B,T)$ = $R^T_{\eta,p}(B,T) R^{D}_{\eta,p}(B) R^{MB}_{\eta,p}(B) R^{s}_{\eta,p}$
where:

\begin{eqnarray}
R^{T}_{\eta,p}= pX_{\eta} \sinh^{-1}(pX_{\eta}),\\
R^{D}_{\eta,p} = \exp(-pu_0m_{\eta}T_DB^{-1}),\\
R^{MB}_{\eta,p} = (ip_0)^{n^t_{\eta}}(q_0)^{n^r_{\eta}},\\
R^{s}_{\eta,p} = \cos(\pi g_{\eta}m_{\eta}/2).
\end{eqnarray}

The field-and temperature-dependent variable ($X_{\eta}$) and the constant ($u_0$) are expressed
as $X_{\eta}$ = $u_0 m_{\eta} T/B$ and $u_0$ = 2$\pi^2 k_B m_e(e\hbar)^{-1}$ = 14.694 T/K. The tunneling ($p_0$) and reflection ($q_0$) probabilities are given by $p_0$ = $e^{-B_0/2B}$
and $p_0^2$ + $q_0^2$ = 1 \cite{Sh84}. $\phi_0=h/e$ is the magnetic flux quantum, $T_D$ is the Dingle temperature
defined by $T_{D}$ = $\hbar(2\pi k_B\tau)^{-1}$, where $\tau^{-1}$ is the
scattering rate, $B_0$ is the MB field, $m_{\eta}$ and $g_{\eta}$ are the
effective masses and effective Land\'{e} factor, respectively. In the case where the magnetic field direction is not parallel to the normal to the conducting plane (angle $\theta$), $B$ is changed to $B\cos\theta$ and the spin damping factor is written $R^{s}_{\eta,p} = \cos(\pi g_{\eta}m_{\eta}/2\cos\theta)$. For
convenience, energies ($E$) such as $\mu$, $\Delta_0$, $\Delta_1$ are expressed in units of
Tesla, using the conversion $(u_0/k_B)[T/J]\times  E[J]= E[T]$.
$B$ and $T$ are the magnetic field [T] and temperature [K], respectively.
Effective masses are expressed in units of the electron mass $m_e$, and magnetization
in Tesla units. The advantage of taking this convention is that field,
frequencies and temperature are not expressed in reduced units.

Frequencies $F_{\eta}$[T] can be written as
$F_{\eta}=m_{\eta}(\mu-\Delta_{\eta})$ and are dependent on the
chemical potential $\mu$ since they are proportional to the area enclosed by
the orbits. Coefficients $C_{\eta}$ are the
symmetry factors of orbits $\eta$. Namely,
$C_{\alpha}=C_{\beta}=C_{2\beta-\alpha}=1$ and $C_{\alpha+\beta}=C_{2\beta}=2$.
Integers $n^t_{\eta}$ and $n^r_{\eta}$ are the number of MB-induced tunnelings and reflections, respectively.
$\varphi_{\eta}$ is the Onsager phase factor of the orbit $\eta$, defined by
the number of turning points, $i.e.$ $\pi/2$ times the number of extrema of the
orbit along one direction (see Fig.~\ref{fig:FS_orbits}). $N$ is given by
$d\Omega/d\mu=-N$, and the chemical potential
satisfies the following implicit equation:

\begin{eqnarray}
\label{mu_0}
\nonumber
\mu=\mu_0-\frac{B}{m_{\beta}}\sum_{p\ge 1}\sum_{\eta}\frac{1}{\pi
p}C_{\eta}
R_{\eta,p}(T)\sin(2\pi p\frac{F_{\eta}}{B}+p\varphi_{\eta}),
\end{eqnarray}

which can be rewritten as

\begin{eqnarray}
\label{mu}
\mu = \mu_0-\frac{B}{m_{\beta}}\sum_{\eta}M_{\eta}(B).
\end{eqnarray}

where $\mu_0$ is the zero-field Fermi energy. For a compensated system, in
which case $N=0$, it is equal to
$\mu_0=(m_0\Delta_0+m_{\alpha}\Delta_{\alpha})/(m_0+m_{\alpha})$. The
oscillatory part of the magnetization is defined as

\begin{eqnarray}
\label{Eq:m_osc}
m_{osc}[T]=-\frac{\phi_0 u_0}{{\cal A}k_B}\frac{\partial F(T,N,B)}{\partial B}.
\end{eqnarray}

In this expression, the free energy, after factorization and simplification, is
given by

\begin{eqnarray}\label{free}
\frac{\phi_0 u_0}{{\cal A}k_B}
F(T,N,B)&=&-\frac{B^2}{2m_{\beta}}\left
(\sum_{\eta}M_{\eta}(B)\right )^2
\\ \nn
&+&\frac{B^2}{2}\sum_{p}\sum_{\eta}\frac{C_{\eta}R_{\eta,p}}{\pi^2p^2m_{\eta}}
\cos\left (2\pi p\frac{F_{\eta}}{B}-2\pi
p\frac{m_{\eta}}{m_{\beta}}\sum_{\eta'}M_{\eta'}(B)
+p\varphi_{\eta}\right )+{\rm cst}
\end{eqnarray}

Since oscillating factors $M_{\eta}$ entering Eq.~\ref{mu} are at first order in
damping factors $R_{\eta,p}(B,T)$ and small compared to $\mu_0$, Eq.
\ref{Eq:m_osc} is solved at second order in $R_{\eta,p}(B,T)$ (the first order
part corresponding to the Lifshitz-Kosevich (LK) semi-classical result). This
leads, after some algebra, to an expansion in power terms of the amplitudes

\begin{eqnarray}\nonumber
m_{osc}&=& -\sum_{\eta}\sum_{p\ge 1}\frac{F_{\eta}C_{\eta}}{\pi p
m_{\eta}}
R_{\eta,p}(B,T)\sin\left ( 2\pi p \frac{F_{\eta}}{B}+p\varphi_{\eta} \right )
\\ \nonumber
&+&\sum_{\eta,\eta'}\sum_{p,p'\ge 1}\frac{F_{\eta}C_{\eta}C_{\eta'}}{\pi p' m_{\beta}}
R_{\eta,p}(B,T)R_{\eta',p'}(B,T)
\left [
\sin\left ( 2\pi \frac{pF_{\eta}+p'F_{\eta'}}{B}+p\varphi_{\eta}+p'\varphi_{\eta'} \right )
\right .
\\ \label{Eq:m_osc2}
&-&\left .\sin\left ( 2\pi \frac{pF_{\eta}-p'F_{\eta'}}{B}+p\varphi_{\eta}-p'\varphi_{\eta'} \right )
\right ]+\cdots
\end{eqnarray}

where the next terms are third order. From this step onwards, frequencies $F_{\eta}$ are evaluated
at $\mu=\mu_0$: $F_{\eta}=m_{\eta}(\mu_0-\Delta_{\eta})$. According to the above
expression, magnetization spectrum can now be expressed in terms of both
classical and non-classical frequencies, still noted as $F_{\eta}$ in the
following, and can be expanded as:

\begin{eqnarray}%\nonumber
\label{Eq:m_osc3}
m_{osc}=\sum_{\eta,p\ge 1}A_{p\eta}\sin\left (2\pi p \frac{F_{\eta}}{B}+p\phi_{\eta} \right ).
\end{eqnarray}

It is important to stress that the amplitude $A_{p\eta}$ involves not only the contribution of
the $p^{th}$ harmonics of the $\eta$ classical orbit, given by the LK formalism
($A_{p\eta}\propto R_{\eta,p}) $ but also higher order corrections,
calculated here at the second order in damping factors. The expressions of the
dominant Fourier components, considered for the data analysis, are given below:\footnote{In Ref. \cite{Au12}, all the Onsager phases are arbitrarily set as 0. For this reason, amplitudes in Eqs. \ref{Eq:alpha}, \ref{Eq:beta} and \ref{Eq:2beta-alpha} have an opposite sign compared to the data in Ref.~\cite{Au12}, accounting for the $\pi$ dephasing reported in Table \ref{tab:table1}.}

\begin{eqnarray}
\label{Eq:alpha}
A_{\alpha}&=&-\frac{F_{\alpha}}{\pi m_{\alpha}}R_{\alpha,1}-
\frac{F_{\alpha}}{\pi m_{\beta}}
\left [
\ff R_{\alpha,1}R_{\alpha,2}
+\frac{1}{6}R_{\alpha,2}R_{\alpha,3}+2R_{\beta,1}R_{\alpha+\beta,1}
+\ff R_{\beta,2}R_{2\beta-\alpha,1}
\right ]
\\ %\nn
\label{Eq:2alpha}
A_{2\alpha}&=&-\frac{F_{\alpha}}{2\pi m_{\alpha}}R_{\alpha,2}+
\frac{F_{\alpha}}{\pi m_{\beta}}\left [
R_{\alpha,1}^2-\frac{2}{3}R_{\alpha,1}R_{\alpha,3}-R_{\alpha,2}
R_{\alpha+\beta,2} \right ]
\\ %\nn
\label{Eq:beta}
A_{\beta}&=&-\frac{F_{\beta}}{\pi m_{\beta}}R_{\beta,1}-
\frac{F_{\beta}}{\pi m_{\beta}}
\left [
\ff R_{\beta,1}R_{\beta,2}
+\frac{1}{6}R_{\beta,2}R_{\beta,3}+2R_{\alpha,1}R_{\alpha+\beta,1}
+2R_{\beta,1}R_{2\beta,1}
\right ]
\\ \nn
\label{Eq:2beta}
A_{2\beta}&=&-\frac{F_{\beta}}{2\pi m_{\beta}}\left
[R_{\beta,2}+2R_{2\beta,1}\right ]+
\frac{F_{\beta}}{\pi m_{\beta}}
\left [
R_{\beta,1}^2-\frac{2}{3}R_{\beta,1}R_{\beta,3}-\frac{1}{4}R_{\beta,2}R_{\beta,4
}-R_{\alpha,2}R_{\alpha+\beta,2}\right .
\\ %\nn
&+&\left .2R_{\alpha,1}R_{2\beta-\alpha,1}-R_{\beta,2}R_{
2\beta,2}-R_{\beta,4}R_{2\beta,1}
\right ]
\\ %\nn
\label{Eq:beta-alpha}
A_{\beta-\alpha}&=&-\frac{F_{\beta-\alpha}}{\pi m_{\beta}}\left [
R_{\alpha,1}R_{\beta,1}+R_{\alpha,2}R_{\alpha+\beta,1}+R_{\beta,2}R_{
\alpha+\beta,1}+R_{\beta,1}R_{2\beta-\alpha,1}\right
]
\\ %\nn
\label{Eq:beta+alpha}
A_{\beta+\alpha}&=&-\frac{2F_{\beta+\alpha}}{\pi m_{\beta+\alpha}}
R_{\beta+\alpha,1}+\frac{F_{\beta+\alpha}}{\pi
m_{\beta}}\left [R_{\alpha,1}R_{\beta,1}-2R_{\alpha+\beta,2}R_{\alpha+\beta,1}
-\frac { 1}{3}R_ { \beta,3}R_{2\beta-\alpha,1} \right
]
\\
\label{Eq:2beta-alpha}
A_{2\beta-\alpha}&=&-\frac{F_{2\beta-\alpha}}{\pi m_{2\beta-\alpha}}
R_{2\beta-\alpha,1}-
\frac{F_{2\beta-\alpha}}{\pi m_{\beta}}
\left [\ff R_{\alpha,1}R_{\beta,2}+
\frac{1}{3}
R_{\alpha,3}R_{\alpha+\beta,2}\right ]
\\ \nn
\label{Eq:2beta-2alpha}
A_{2\beta-2\alpha}&=&-\frac{F_{2\beta-2\alpha}}{\pi m_{\beta}}
\left [2R_{\alpha,2}R_{2\beta,1}+2R_{2\beta-\alpha,2}R_{2\beta,1}
+2R_{\alpha,1}R_{2\beta-\alpha,1}+\frac{1}{2}R_{\alpha,4}R_{\alpha+\beta,2}
\right .
\\
&+&\left .
\frac{1}{2}R_{\alpha,2}R_{\beta,2}+\frac{1}{2}R_{\beta,2}R_{2\beta-\alpha,2}
\right ].
\end{eqnarray}

As discussed in Ref.~\cite{Au12}, the leading term of Eqs.~\ref{Eq:alpha}, \ref{Eq:beta} and \ref{Eq:2beta-alpha},
relevant to the classical orbits, $\alpha$, $\beta$ and $2\beta-\alpha$, respectively, corresponds to the LK formalism.
This statement also holds for $\beta+\alpha$, $2\alpha$ and $2\beta$ even though the second order terms may have
magnitude close to the first order term, being able to yield non-monotonous field and temperature dependence.
In the specific case of Eq.~\ref{Eq:2beta}, the leading term involves the contributions of both the classical
orbit $2\beta$ displayed in Fig.~\ref{fig:FS_orbits} and the second harmonics of $\beta$ which are accounted for by
the damping factors $R_{2\beta,1}$ and $R_{\beta,2}$, respectively. In contrast, there is no first order term
entering Eqs.~\ref{Eq:beta-alpha} and \ref{Eq:2beta-2alpha} relevant to $\beta-\alpha$ and its second harmonics,
respectively, since these Fourier components correspond to 'forbidden frequencies'.

\subsection{Onsager phase factor}

\begin{figure}
\centering
%\resizebox{\columnwidth}{!}
\resizebox{12cm}{!}{\includegraphics*{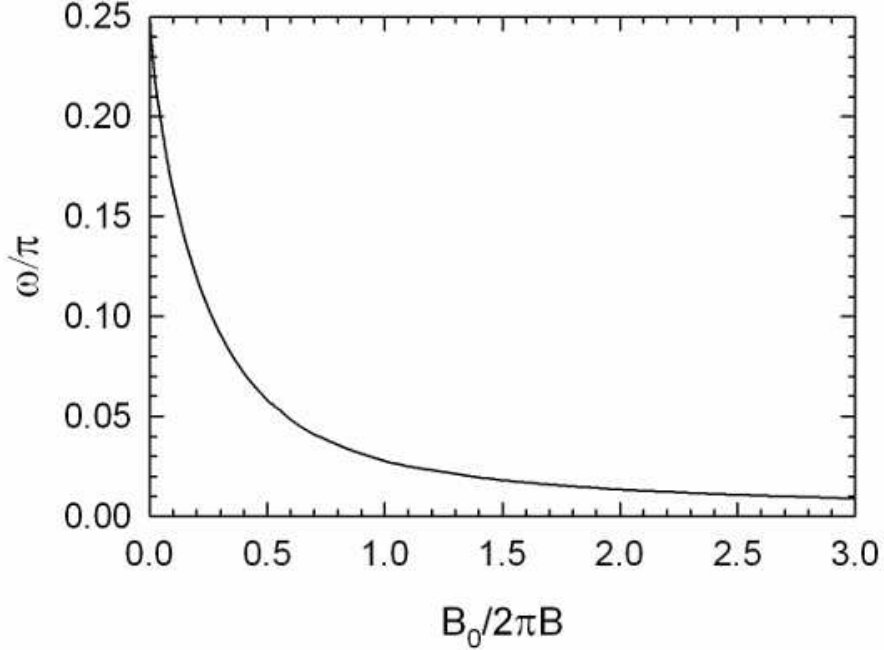}}
\caption{\label{fig:omega} Phase $\omega$ as a function of the argument
$B_0/(2\pi B)$. For small fields $B$ compared to the magnetic breakdown field
$B_0$, the phase vanishes while it goes to $\pi/4$ at large fields.}
\end{figure}

Turn on now to the determination of the Onsager phase factors $\phi_{\eta}$ entering Eq.~\ref{Eq:m_osc3}. Besides a phase factor  $\varphi_{\eta}$, deduced from the classical orbits' phases appearing in Eq.~\ref{eq:omega}, an additional phase factor $\omega$ is added to $\varphi_{\eta}$ each time a quasiparticle is reflected at a MB junction. Indeed, according to Refs. \cite{Sl67,Ko68,Hu76}, the matrix for the
incoming and outgoing wave-function amplitudes at each junction point is given by

\begin{equation}
\label{Eq:matrix}
M =
\left (
\begin{array}{cc}
q_0{\rm e}^{-i\omega} & ip_0 \\
ip_0 & q_0{\rm e}^{i\omega} \\
\end{array}
\right )
\end{equation}

with

\begin{equation}
\omega(B)=-\frac{\pi}{4}+x\log(x)-x-\arg \Gamma(ix),\;\;
x=\frac{B_0}{2\pi B}.
\end{equation}

After a reflection, the quasi-particle amplitude takes a factor
$q_0\exp(-i\omega)$ and $q_0\exp(i\omega)$ for quasi-particle path orientation clockwise and counter-clockwise, respectively.
As displayed in  Fig. \ref{fig:omega}, $\omega$ goes to zero at low field. In contrast, it takes noticeable
values as the field is larger than $B_0$, going to $\pi$/4 at large field.
For example, for $B_0$ = 35 T, and $B$ = 55 T, that are relevant values for the
compound studied in this paper, we obtain $\omega$ = 0.16$\pi$, which is
not negligible, especially if the number of reflections events  $n^r_{\eta}$  is large. According to Eq.~\ref{Eq:matrix}, the Onsager phase factor is given by

\begin{equation}
\label{Eq:phi}
\phi_{\eta} = \varphi_{\eta} - n^r_{\eta}\omega(B)
\end{equation}

The $\varphi_{\eta}$ and
$n^r_{\eta}$ values relevant to the Fourier components appearing in
Eqs.~\ref{Eq:alpha} to \ref{Eq:2beta-2alpha} are given in Table~\ref{tab:table1}.
We notice in particular that the index $n^r_{\eta}$ can be negative, due
to algebraic combinations of the individual phases present in the sine function
of Eq.~\ref{Eq:m_osc2}. It can also be remarked that the Fourier component with frequency $F_{2\beta}$ arises from the second harmonics of $\beta$ and the 2$\beta$ orbit displayed in Fig.~\ref{fig:FS_orbits}. Nevertheless, these two contributions have the same Onsager phase. Besides, for a given $\eta$ Fourier component, all the involved second order terms (see Eqs.~\ref{Eq:alpha} to \ref{Eq:2beta-2alpha}) can be viewed as arising from algebraic combinations of classical orbits yielding the same Onsager phase.

\begin{table}[h]
\caption{\label{tab:table1}
Onsager phase factors predicted by Eq.~\ref{Eq:phi} for the various Fourier components considered for the data analysis and $\varphi_{\eta}/\pi$ values accounting for the best fits of Eqs.~\ref{Eq:alpha} to~\ref{Eq:2beta-2alpha} to the oscillatory torque data in the temperature range 1.4 - 4.2 K (see Fig.~\ref{fig:fit} for the data at 1.4 K), assuming either constant (assumption (i): $n^r_{\eta}$ = 0 in Eq.~\ref{Eq:phi}) or field-dependent Onsager phase factor (assumption (ii)).
}
\smallskip
%\begin{ruledtabular}
\begin{tabular}{cccccccc}
\hline
Fourier & \multicolumn{2}{c}{Predicted (Eq.~\ref{Eq:phi}) } &
\multicolumn{2}{c}{fittings $\phi_{\eta}$ = $\varphi_{\eta}$}&
\multicolumn{3}{c}{fittings $\phi_{\eta}$ = $\varphi_{\eta}$ -
$n^r_{\eta}\omega$}\\
component & $\varphi_{\eta}/\pi$ & $n^r_{\eta}$ & F (T) & $\varphi_{\eta}/\pi$ & F (T) & $\varphi_{\eta}/\pi$ \\ % & $n_{i}$
\hline
%\colrule
$\alpha$         & 1 & 2 & 949.7 $\pm$ 1.2& 0.75 $\pm$ 0.04     & 947.3 $\pm$
1.4  & 1.16 $\pm$ 0.05\\
2$\alpha$        & 0 & 4  &                & -0.52 $\pm$ 0.21    &                  & 0.30 $\pm$ 0.13\\
$\beta-\alpha$   & 0 & -2  &                & 0.35 $\pm$ 0.21     &
    & -0.06 $\pm$ 0.21\\
$\beta$          & 1 & 0  & 4631 $\pm$ 6   & 1.08 $\pm$ 0.19     & 4631 $\pm$ 6     & 1.08 $\pm$ 0.19\\
$\beta+\alpha$   & 0 & 2  &                & -0.7 $\pm$ 0.8      &                  & -0.3 $\pm$ 0.8\\
$2\beta-2\alpha$& 0 & -4  &                & 0.5 $\pm$ 0.7       &
    & -0.3 $\pm$ 0.7\\
2$\beta-\alpha$  & 1 & -2  &                & -0.54 $\pm$ 0.34    &
    & 1.05 $\pm$ 0.34\\
2$\beta$         & 1 & 0  &                & 1.1 $\pm$ 0.5       &                  & 1.1 $\pm$ 0.5\\
\hline
\end{tabular}

%\end{ruledtabular}
\end{table}

\section{\label{sec:expt}Experimental}

Field- and temperature-dependent magnetic torque of the considered crystal was studied in Ref. \cite{Au12}. It was synthesized by electrocrystallization technique as reported in Ref.~\cite{Sh11}.  Its size is
approximately 0.12 $\times$ 0.1 $\times$ 0.04~mm$^3$. Recall that magnetic torque was
measured  with a commercial piezoresistive microcantilever, in pulsed magnetic
fields of up to 55 T with a pulse decay duration of 0.32 s. Variations of the
cantilever piezoresistance were measured in the temperature range from 1.4 K to
4.2 K with a Wheatstone bridge with an $ac$ excitation at a frequency of 63 kHz.
The angle between the normal to the conducting plane and the magnetic field
direction was $\theta$ = 7$^{\circ}$.

\section{\label{sec:res}Results and discussion}

\begin{figure}
\centering
%\resizebox{\columnwidth}{!}
\resizebox{12cm}{!}{\includegraphics*{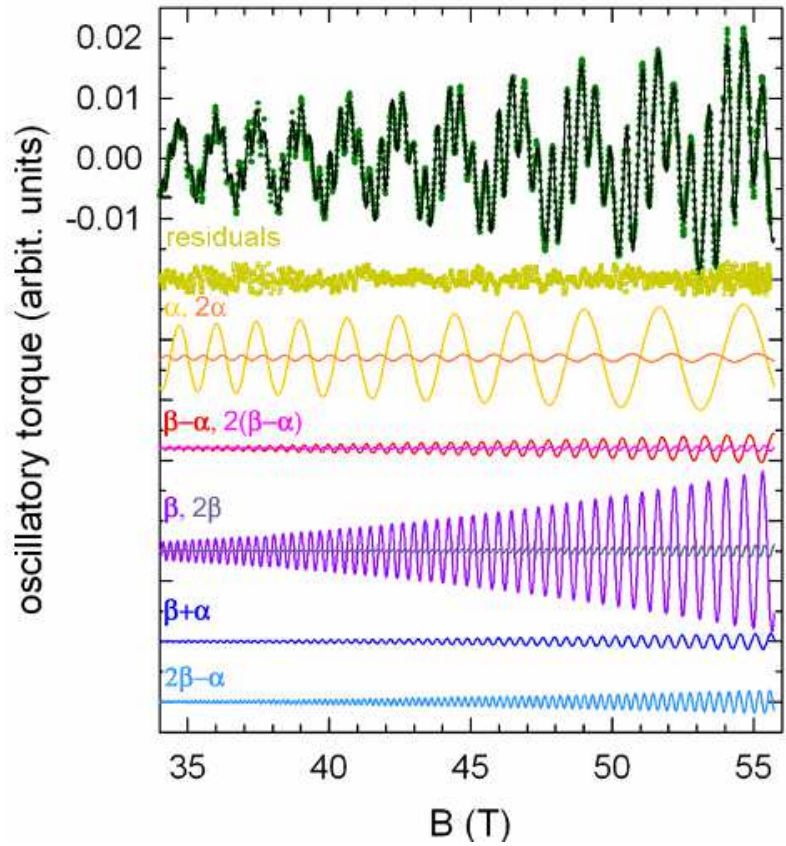}}
\caption{\label{fig:fit} (Color online) Oscillatory torque data at 1.4 K (solid circles) and best fit of Eqs.~\ref{Eq:alpha} to \ref{Eq:2beta-2alpha} (green solid line). The
residuals (open squares) and the various Fourier components entering the fittings (solid lines) are shifted down from
each others.}
\end{figure}

The field-dependent amplitudes A$_{\eta}$ of the various Fourier components
entering the spectra can be analyzed on the basis of Eqs.~\ref{Eq:alpha} to
\ref{Eq:2beta-2alpha}, keeping in mind that they are related to torque
oscillations amplitudes $A^{\tau}_{\eta}$ as $A_{\eta} = \tau_0
A^{\tau}_{\eta}$/($B$tan$\theta$) where $\tau_0$ is a prefactor depending of the cantilever stiffness, crystal mass, $etc.$. Onsager phase factors are considered
within the assumptions of (i) constant ($n^r_{\eta}$ = 0, $i.e.$ $\phi_{\eta}$ =
$\varphi_{\eta}$ in Eq.~\ref{Eq:phi}) and (ii) field-dependent $\phi_{\eta}$.

The field range above 20 T, in which dHvA oscillations are observed is
considered for the data analysis. As discussed above, numerous physical
parameters enter the oscillatory spectra. In order to reduce the number of free
parameters for the fittings, effective masses ($m_{\alpha}$ = 1.81, $m_{\beta}$
= 3.52), MB field ($B_0$ = 35 T) and Dingle temperature ($T_D$ = 0.79 K for all
the classical orbits) are taken from Ref.~\cite{Au12} and kept fixed.
Besides, the effective Land\'{e} factor is assumed to be the same for all the
orbits ($g_{\eta}$ = $g^*$). As a result, besides prefactors $\tau_0$ and the Land\'{e} factor  $g^*$, only the two frequencies $F_{\alpha}$ and  $F_{\beta}$ and the various field-independent parts of the Onsager phase factors $\varphi_{\eta}$ are free parameters.

As an example, the high field range of the data measured at 1.4 K are reported
in Fig.~\ref{fig:fit}. Best fits obtained within either assumptions (i) or (ii)
are indiscernible to the naked eye in the field range considered for the data analysis, even though, strictly speaking, oscillations are no more periodic in 1$/B$ within
Eq.~\ref{Eq:phi}. Actually, it can be checked that, due to the limited field
range in which oscillations are observed, Fourier transforms of the best fits
obtained within either the assumption of (i) constant or (ii) field-dependent Onsager
phase, are indiscernible, as well. The deduced effective Land\'{e} factor is
$g^*$ = 1.9 $\pm$ 0.2, in agreement with the reported value of
Ref.~\cite{Au12}. Deduced values of $\varphi_{\eta}$, $F_{\alpha}$ and
$F_{\beta}$ are given in Table~\ref{tab:table1}.

As expected, $F_{\beta}$, $\varphi_{\beta}$ and $\varphi_{2\beta}$ are insensitive to the considered assumption (i) or (ii) since only tunnelings enter $\beta$ and 2$\beta$ orbits ($n_{\beta}$ = 0 and $n_{2\beta}$ = 0 in Eq.~\ref{Eq:phi}). In addition, $\varphi_{\beta}$ is in agreement with the predicted value within the error bars. This is also the case of $\varphi_{2\beta}$ although a large uncertainty is obtained due to the small amplitude of this Fourier component (see Fig.~\ref{fig:fit}).

In contrast, the value of both the frequency $F_{\alpha}$ and Onsager phases of
the Fourier components involving $\alpha$ depend on the considered assumption
(i) or (ii). Crudely speaking, the $F_{\alpha}$ value deduced from fittings within assumption (ii) accommodates to compensate the
field-dependent phase. However, the observed change is small and remains within
the uncertainty. Not any of these Onsager phase values, deduced within assumption (i),
are in agreement with the predictions (see Table~\ref{tab:table1}). In other words, field-independent Onsager phases given by $\pi$/2 times the number of orbit extrema in $k$-space cannot account for the data.  Assuming a
field-dependent phase factor, $\varphi_{\alpha}$ and $\varphi_{2\alpha}$ come much
closer to the predictions of the model, even though the discrepancy with the
predicted values are still slightly off the error bars. Besides,
$\varphi_{\beta-\alpha}$ and $\varphi_{2\beta-\alpha}$ are in very good
agreement with the predictions. This statement also stands for
$\varphi_{\beta+\alpha}$ and $\varphi_{2\beta-2\alpha}$ albeit the error bars
are very large due to the small amplitude of these Fourier components.

\section{\label{sec:conclusion}Conclusion}

Many years ago, theoretical calculations predicted that Onsager phase factor of quantum oscillations includes a field-dependent part in the case of a magnetic breakdown orbit with reflections \cite{Sl67,Ko68,Hu76}. To our best knowledge, this feature was only considered in the case of Cd \cite{Co71} which is a three-dimensional elemental metal.

It is demonstrated that field-dependent phase is necessary to account for the oscillatory spectrum of the two-dimensional organic metal $\theta$-(BEDT-TTF)$_4$CoBr$_4$(C$_6$H$_4$Cl$_2$). The Fermi surface of this compound achieves the network of coupled orbits model proposed by Pippard more than fifty years ago\cite{Pi62} which is relevant for many organic compounds. In agreement with the above mentioned theoretical predictions, magnetic breakdown orbits involving reflections, namely all the orbits including the $\alpha$ component for the considered Fermi surface topology, are accounted for by field-dependent Onsager phase factors. This result confirms that field-dependent phase factor is a general feature of magnetic breakdown orbits.

As a result, the value of $F_{\alpha}$ deduced within Eq.~\ref{Eq:phi} is slightly reduced compared to the value derived assuming constant Onsager phases (i.e. through either direct fitting assuming $\phi_{\eta}$ = $\varphi_{\eta}$ or Fourier analysis of the data), even though the observed discrepancy stays within the error bars. Despite field-dependent phase factors, the magnetic oscillations periodicity in 1/$B$ is preserved, within the
experimental uncertainty. This behaviour is likely due to the relatively small field range considered (20 T to 55 T). Obviously, keeping in mind that the magnetic breakdown field of the studied compound is rather large ($B_0$ = 35 T), larger effects are expected for compounds with smaller magnetic breakdown field.

\section*{Acknowledgements}
This work has been supported by EuroMagNET II under the EU contract number 228043 and by the CNRS-RFBR cooperation under the PICS contract number 5708. Work at Bellaterra was supported by MINECO (Grants Projects FIS2009-1271-C04-03 and CSD 2007-00041).



%% References


%\bibliographystyle{model1b-num-names}

%% not want to use model1b-num-names.bst.


\end{document}